\documentclass{llncs}

\usepackage[T1]{fontenc}
\usepackage{amssymb}

\def\la{\leftarrow}
\def\ra{\rightarrow}
\def\ex{\exists}
\def\fa{\forall}

\newtheorem{ntheorem}{Theorem}

\newenvironment{nproof}{{\noindent {\it Proof.}~}}{\medskip}

\newtheorem{fdefinition}{Définition}
\newtheorem{ftheorem}{Théorème}
\newtheorem{fproposition}{Proposition}
\newenvironment{preuve}{{\noindent {\it Preuve.}~}}{\medskip}

\newbox\tempa
\newbox\tempb
\newdimen\tempc
\def\mud#1{\hfil $\displaystyle{\mathstrut #1}$\hfil}
\def\rig#1{\hfil $\displaystyle{#1}$}
\def\irulehelp#1#2#3{\setbox\tempa=\hbox{$\displaystyle{\mathstrut #2}$}%
		        \setbox\tempb=\vbox{\halign{##\cr
	\mud{#1}\cr
	\noalign{\vskip\the\lineskip}%
	\noalign{\hrule height 0pt}%
	\rig{\vbox to 0pt{\vss\hbox to 0pt{${\; #3}$\hss}\vss}}\cr
	\noalign{\hrule}%
	\noalign{\vskip\the\lineskip}%
	\mud{\copy\tempa}\cr}}%
		      \tempc=\wd\tempb
		      \advance\tempc by \wd\tempa
		      \divide\tempc by 2 }
\def\irule#1#2#3{{\irulehelp{#1}{#2}{#3}%
		     \hbox to \wd\tempa{\hss \box\tempb \hss}}}

\begin{document}
\begin{center}{\Large{\bf L'Indécidabilité du Filtrage du Troisième Ordre 
dans les Calculs avec Types Dépendants ou Constructeurs de Types}}\\[20pt]
{\bf Gilles Dowek}\\[10pt]
\end{center}
\thispagestyle{empty}

\noindent {\bf Résumé}~: On prouve l'indécidabilité du problème du
filtrage du troisième ordre dans les $\lambda$-calcul typés avec types
dépendants et dans ceux avec constructeurs de types en leur réduisant
le problème de l'unification du second ordre.

\begin{center}{\Large{\bf The Undecidability of Third Order Pattern Matching
in Calculi with  Dependent Types or Type Constructors}}\\[20pt]
\end{center}

\noindent {\bf Summary}: 
We prove the undecidability of the third order pattern matching problem
in typed $\lambda$-calculi with dependent types and in those with 
type constructors by reducing the second order unification problem to them.

\section*{Abridged English Version}

We define eight typed $\lambda$-calculi, called the {\it calculi of the cube} 
\cite{Barendregt}. All these calculi allow functions from terms to terms. 
Those that allow functions from terms to types are said to admit 
{\it dependent types},  those that allow functions from types to terms are said
to be {\it polymorphic}, those that allow functions from types to types are 
said to admit {\it types constructors}.
We write $Prop$ and $Type$ for the sorts written $*$ and $\Box$ in
\cite{Barendregt}.

A {\it unification problem} is a triple $\langle \Gamma,a,b\rangle $ such that 
$\Gamma$ is a context of universally and existentially quantified variables
and $a$ and $b$ are well-typed terms of the same type in $\Gamma$.
A {\it matching problem} is a unification problem such that $b$ is closed in
$\Gamma$.
A substitution $\sigma$ is said to be a {\it solution} of the problem 
$\langle \Gamma,a,b\rangle $ if it is well-typed in $\Gamma$ and the terms $\sigma a$ and 
$\sigma b$ are equivalent.

A unification problem $\langle \Gamma,a,b\rangle $ is said to be {\it term-elementary} if and
only if $\Gamma = [\fa U:Prop]{\Gamma}'$, for all declarations
$Qx:T$ of ${\Gamma}'$, the term $T$ is one of the terms
$U$, $U \ra U$, $U \ra U \ra U$ or $U \ra U \ra U \ra U$
and the common type of $a$ and $b$ is $U$.
In a calculus with type constructors, a unification problem $\langle \Gamma,a,b\rangle $
is said to be {\it type-elementary}, if and only if
for all declarations $Qx:T$ of $\Gamma$, the term $T$ is one of the terms
$Prop$, $Prop \ra Prop$, $Prop \ra Prop \ra Prop$ or
$Prop \ra Prop \ra Prop \ra Prop$ and the common type
of $a$ and $b$ is $Prop$.

Goldfarb's theorem \cite{Goldfarb} generalizes to all the calculi of the cube
and shows that in all these calculi, there is no effective method that 
decides if a term-elementary unification problem has a solution and in the
calculi with type constructors there is no effective method that decides if 
a type-elementary unification problem has a solution.

\begin{ntheorem}
In a calculus with dependent types there is no effective method that decides 
if a matching problem in which all existential variables have at most 
third order types has a solution.
\end{ntheorem}

\nproof{For every term-elementary unification problem
$\langle \gamma, u_{1}, u_{2}\rangle $, we construct 
a matching problem $\langle \Gamma,t_{1},t_{2}\rangle $ such that all the types of 
the existential variables of $\Gamma$ are at most third order and
$\langle \gamma,u_{1},u_{2}\rangle $ has a solution if and only if $\langle \Gamma,t_{1},t_{2}\rangle $ 
also has one.
We let:
$$\Gamma = \gamma[\fa z:U; \fa P:U \ra Prop; \fa c:(P~z); \fa d:(P~z);
\fa G:(P~z) \ra (P~z) \ra (P~z);$$
$$\ex f:(h:U \ra U)(P~(h~(u_{1}))) \ra (P~(h~(u_{2})))]$$
$$t_{1} = (G~(f~[x:U]z~c)~(f~[x:U]z~d))~~~~~~~~t_{2} = (G~c~d)$$

If there exists a substitution $\tau$ such that 
$\tau u_{1} = \tau u_{2}$ then we let:
$$\sigma =\tau \cup 
\{\langle f,[],[x_{1}:U \ra U][x_{2}:(P~(x_{1}~(\tau u_{1})))]x_{2}\rangle \}$$
The substitution $\sigma$ is well-typed in $\Gamma$ and is a solution for the
matching problem $\sigma t_{1} = t_{2}$.

Conversely if there exists a substitution $\sigma$ well-typed in 
$\Gamma$ such that
$\sigma t_{1} = t_{2}$, then $\sigma$ is well-typed in $\gamma$ and
$\sigma f$ must be the term 
$[x_{1}:U \ra U][x_{2}:(P~(x_{1}~(\sigma u_{1})))]x_{2}$,
so $\sigma u_{1} = \sigma u_{2}$.}

\begin{ntheorem}
In a calculus with type constructors there is no effective method that
decides if a matching problem in which all existential variables have at most 
third order types has a solution.
\end{ntheorem}

\nproof{For every type-elementary unification problem
$\langle \gamma, u_{1}, u_{2}\rangle $ we build a matching problem
$\langle \Gamma,t_{1},t_{2}\rangle $ 
such that all the types of existential variables of $\Gamma$ 
are at most third order and 
$\langle \gamma, u_{1}, u_{2}\rangle $ has a solution if and only if 
$\langle \Gamma,t_{1},t_{2}\rangle $ also has one.
We let:
$$\Gamma = \gamma[\fa Z:Prop; \fa c:Z; \fa d:Z; \fa G:Z \ra Z \ra Z;
\ex f:(h:Prop \ra Prop)(h~u_{1}) \ra (h~u_{2})]$$
$$t_{1} = (G~(f~[X:Prop]Z~c)~(f~[X:Prop]Z~d))~~~~~~~~t_{2} = (G~c~d)$$

If there exists a substitution $\tau$ such that
$\tau u_{1} = \tau u_{2}$ then we let:
$$\sigma =\tau \cup 
\{\langle f,[],[x_{1}:Prop \ra Prop][x_{2}:(x_{1}~(\tau u_{1}))]x_{2}\rangle \}$$
The substitution $\sigma$ is well-typed in $\Gamma$ and is a solution of the
matching problem $\sigma t_{1} = t_{2}$.

Conversely if there exists a substitution $\sigma$ well-typed in 
$\Gamma$ such that 
$\sigma t_{1} = t_{2}$, then $\sigma$ is well typed in $\gamma$ and
$\sigma f$ must be the term 
$[x_{1}:Prop \ra Prop][x_{2}:(x_{1}~(\sigma u_{1}))]x_{2}$ 
so $\sigma u_{1} = \sigma u_{2}$.}

\section{Le Cube des $\lambda$-calculs Typés}

On définit huit $\lambda$-calculs typés appelés
les {\it calculs du cube} \cite{Barendregt}.
Tous ces calculs autorisent les fonctions des termes dans les termes.
Ceux qui autorisent les fonctions des termes dans les types sont dits 
admettre des {\it types dépendants}, ceux qui autorisent des fonctions des 
types dans les termes sont dits {\it polymorphes}, ceux qui autorisent des 
fonctions des types dans les types sont dits admettre des 
{\it constructeurs de types}. Un calcul peut avoir ou non chacune de ces trois 
propriétés.

\begin{fdefinition}[Syntaxe]
$$T~::=~Prop~|~Type~|~x~|~(T~T)~|~[x:T]T~|~(x:T)T$$
Dans cette note, on ignore les problèmes de capture de variables. Une
présentation rigoureuse utiliserait des indices de de Bruijn.
Un terme de la forme $[x:T]T'$ est appelé une $\lambda$-abstraction et un
terme de la forme $(x:T)T'$ est appelé un produit.
La notation $T \ra T'$ est utilisée pour $(x:T)T'$ quand $x$ n'a pas
d'occurrence dans $T'$.

Soient $t$ et $t'$ deux termes et $x$ une variable, on note $t[x \la t']$ le 
terme obtenu en substituant le terme $t'$ à la variable $x$ dans le terme $t$.
On note $t = t'$ la relation de $\beta \eta$-équivalence c'est-à-dire la 
plus petite relation 
d'équivalence compatible avec la structure de terme qui vérifie :
$$(([x:T]t) u) =  t[x \la u]~~~(\beta)$$ 
$$[x:T](t~x) = t~ \mbox{si $x$ n'est pas libre dans $t$}~~~(\eta)$$ 
\end{fdefinition}

\begin{fdefinition}
Un {\it contexte} est une liste de couples $x:T$ où $x$ est une variable et 
$T$ un terme.
\end{fdefinition}

\begin{fdefinition}
Soit $R \subseteq \{\langle Prop,Prop\rangle ,\langle Prop,Type\rangle ,\langle Type,Prop\rangle ,\langle Type,Type\rangle \}$ 
tel que $\langle Prop,Prop\rangle  \in R$. On définit deux jugements : 
{\it $\Gamma$ est bien-formé} et 
{\it $t$ a le type $T$ dans $\Gamma$} ($\Gamma \vdash t:T$) où $\Gamma$ est
un contexte et $t$ et $T$ deux termes :
$$\irule{}
         {[]~\mbox{bien-formé}}
         {}$$
$$\irule{\Gamma \vdash T:s}
         {\Gamma[x:T]~\mbox{bien-formé}}
         {s \in \{Prop, Type\}}
$$
$$\irule{\Gamma~\mbox{bien-formé}}
         {\Gamma \vdash Prop:Type}
         {}
$$
$$\irule{\Gamma~\mbox{bien-formé}~~x:T \in \Gamma} 
         {\Gamma \vdash x:T}
         {}$$
$$\irule{\Gamma \vdash T:s~~\Gamma[x:T] \vdash T':s'}
         {\Gamma \vdash (x:T)T':s'}
         {\langle s,s'\rangle  \in R}$$
$$\irule{\Gamma \vdash (x:T)T':s~~\Gamma[x:T] \vdash t:T'}
         {\Gamma \vdash [x:T]t:(x:T)T'}
         {s \in \{Prop,Type\}}$$
$$\irule{\Gamma \vdash t:(x:T)T'~~\Gamma \vdash t':T}
         {\Gamma \vdash (t~t'):T'[x \la t']}
         {}$$
$$\irule{\Gamma \vdash T:s~~\Gamma \vdash T':s~~\Gamma \vdash t:T~~T = T'}
         {\Gamma \vdash t:T'}
         {s \in \{Prop,Type\}}$$
Le calcul tel que $R = \{\langle Prop,Prop\rangle \}$ est appelé {\it $\lambda$-calcul
simplement typé}. Les calculs tels que $\langle Prop,Type\rangle  \in R$ sont
dits admettre des {\it types dépendants}, ceux tels que $\langle Type,Prop\rangle  \in R$ 
sont dits {\it polymorphes}, ceux tels que $\langle Type,Type\rangle  \in R$ 
sont dits admettre des {\it constructeurs de types}.
\end{fdefinition}

\begin{fdefinition}
Soient $t$ et $t'$ deux termes,
la relation de {\it $\beta \eta$-réduction} $t \rhd t'$ est la plus petite
relation réflexive, transitive et compatible avec la structure de terme
qui vérifie :
$$(([x:T]t) u) \rhd t[x \la u]~~~(\beta)$$
$$[x:T](t~x) \rhd t~ \mbox{si $x$ n'est pas libre dans $t$}~~~(\eta)$$ 
On conjecture que la relation de réduction sur les termes bien-typés est 
fortement normalisable et Church-Rosser. De ce fait, pour tout terme on peut 
définir une forme normale unique. Deux termes sont équivalents s'ils ont la 
m\^eme forme normale.
\end{fdefinition}

\begin{fproposition}
Un terme normal $t$ est une abstraction, un produit ou un
terme atomique c'est-à-dire un terme de la forme $(x~t_{1}~...~t_{n})$ 
où $x$ est une variable ou l'un des symboles 
$Prop$ et $Type$. 
\end{fproposition}

\section{Contextes Quantifiés}

\begin{fdefinition}
Un {\it Contexte Quantifié} est un contexte $\Gamma$ à chaque variable duquel
on associe un quantificateur ($\fa$ ou $\ex$).
Dans un contexte $\Gamma$, une variable à laquelle on associe le 
quantificateur universel (resp. existentiel) est dite {\it universelle} 
(resp. {\it existentielle}) dans ce contexte.
\end{fdefinition}

\begin{fdefinition}
Un terme $t$ bien-typé dans un context $\Gamma$
est dit {\it fermé dans $\Gamma$} si pour tout $x$ libre dans $t$, 
$x$ est universelle dans $\Gamma$ et son type est fermé dans le  préfixe de 
$\Gamma$ défini avant $x$.
\end{fdefinition}

\begin{fdefinition}
Un ensemble fini $\sigma$ de triplets $\langle x,\gamma,t\rangle $ où $x$ est une variable 
$\gamma$ un contexte dont toutes les variables sont existentielles
et $t$ est un terme est appelé une {\it substitution} si pour toute variable
$x$ il existe au plus un triplet de la forme $\langle x,\gamma,t\rangle $ dans $\sigma$. 
\end{fdefinition}

\begin{fdefinition}
Soit $x$ une variable et $\sigma$ une substitution. 
s'il y a  un triplet $\langle x,\gamma,t\rangle $ dans $\sigma$ on pose $\sigma x = t$,
sinon on pose $\sigma x = x$. Cette définition s'étend aux termes de
manière naturelle.
\end{fdefinition}

\begin{fdefinition}
Soit un contexte quantifié $\Gamma$ et une substitution $\sigma$.
On définit, par récurrence sur la longueur de $\Gamma$, une relation de
compatibilité :
{\it $\sigma$ est bien-typée dans $\Gamma$}, et si $\sigma$ est bien-typée
dans $\Gamma$, un contexte quantifié $\sigma \Gamma$.\\
Si $\Gamma = []$ alors $\sigma$ est bien-typée dans $\Gamma$ et
$\sigma \Gamma = []$.\\
Si $\Gamma = \Delta[\fa x:T]$ alors
si $\sigma$ est bien-typée dans $\Delta$ on pose ${\Gamma}' = \sigma \Delta$, 
si ${\Gamma}' \vdash \sigma T:Prop$ ou ${\Gamma}' \vdash \sigma T:Type$ 
alors $\sigma$ est bien-typée dans $\Gamma$ et
$\sigma \Gamma = {\Gamma}'[\fa x:\sigma T]$.
Dans les autres cas $\sigma$ n'est pas bien-typée dans $\Gamma$.\\
Si $\Gamma = \Delta[\ex x:T]$ alors soit 
$\gamma$ le contexte associé à $x$ par $\sigma$ si il existe un triplet
$\langle x,\gamma,t\rangle $ dans $\sigma$ et $\gamma = [\ex x:\sigma T]$ sinon.
Si $\sigma$ est bien-typée dans $\Delta$ on pose ${\Gamma}' = \sigma \Delta$.
Si ${\Gamma}'\gamma$ est bien-formé et 
${\Gamma}'\gamma \vdash \sigma x:\sigma T$
alors $\sigma$ est bien-typée dans $\Gamma$ et 
$\sigma \Gamma = {\Gamma}'\gamma$.
Dans les autres cas $\sigma$ n'est pas bien-typée dans $\Gamma$.
\end{fdefinition}

\begin{fdefinition}
Soit $\Gamma$ un contexte quantifié et $T$ un terme normal de type $Prop$
ou $Type$. Le terme $T$ n'est pas une abstraction.
L'{\it ordre} de $T$ ($o(T)$) est l'élément de $N \cup \{\infty\}$
défini par :

Si $T$ est atomique $T = (x~t_{1}~...~t_{n})$ alors si $x$ est une variable 
universelle de $\Gamma$ et on pose $o(T) = 1$, si $x$ est une variable
existentielle de $\Gamma$ on pose $o(T) = \infty$, si $x = Prop$ on pose
$o(T) = 2$.

Si $T$ est un produit $T = (y:U)V$ alors on pose $o(T) = max \{1 + u, v\}$
où $u$ est l'ordre de $U$ dans le contexte quantifié $\Gamma$ et 
$v$ est l'ordre de $V$ dans le contexte quantifié $\Gamma[\ex y:U]$.
\end{fdefinition}

\begin{fdefinition}
Soit $\Gamma$ un contexte et $a$ et $b$ deux termes bien typés et de m\^eme 
type dans $\Gamma$. Le triplet $\langle \Gamma,a,b\rangle $ est appelé {\it problème 
d'unification}. Si le terme $b$ est fermé dans $\Gamma$ il est appelé {\it
problème de filtrage}. Une substitution $\sigma$ est dite {\it solution} du
problème $\langle \Gamma,a,b\rangle $ si elle est bien typée dans $\Gamma$ et 
les termes $\sigma a$ et $\sigma b$ sont équivalents.
\end{fdefinition}

\section{Preuves d'Indécidabilité}

On remarque tout d'abord que la preuve d'indécidabilité de l'unification 
du second ordre dans le $\lambda$-calcul simplement typé
(Goldfarb \cite{Goldfarb}) est valide dans tous les calculs du cube. 
En effet, en reprenant les notations de \cite{Goldfarb},
si $H$ est un système d'équations arithmétiques et $S$ est le problème 
d'unification codant ce système, comme dans le $\lambda$-calcul simplement 
typé, pour toute solution $\theta$ de $S$, 
les variables $F_{i}$ sont instanciées par des termes de la forme:
$$\theta F_{i} = [w_{1}:U]\overline{n_{i}}~w_{1}$$
et les variables $G_{l}$ pour $l = 2^{i}3^{j}5^{k}$ par des termes de la
forme:
$$\theta G_{l} = [w_{1}:U][w_{2}:U][w_{3}:U]
(g~(t_{0}^{l}~w_{1}~w_{2})~(g~(t_{1}^{l}~w_{1}~w_{2})~...~
(g~(t_{n_{j}-1}^{l}~w_{1}~w_{2})~w_{3})))$$
avec:
$$t_{p}^{l} = [w_{1}:U][w_{2}:U]
(g~(\overline{n_{i}.p}~w_{1})~(\overline{p}~w_{2}))$$
$$\overline{n} = [w_{1}:U](g~a~...~(g~a~w_{1}))~\mbox{($n$ fois)}$$
de tout unificateur de $S$ on peut donc déduire une solution de $H$.

\begin{fproposition}[Goldfarb]
Dans un calcul quelconque du cube 
il n'existe pas de méthode effective permettant de décider si un
problème d'unification du second ordre a une solution.
\end{fproposition}

\medskip

De plus :

\begin{fdefinition}
Un problème d'unification $\langle \Gamma,a,b\rangle $ est dit {\it élémentaire au
niveau des termes} si
$\Gamma = [\fa U:Prop]{\Gamma}'$, pour toute variable $Qx:T$ déclarée dans
${\Gamma}'$, le terme $T$ est l'un des termes
$U$, $U \ra U$, $U \ra U \ra U$ ou $U \ra U \ra U \ra U$ et le type commun de
$a$ et $b$ est $U$.

Dans un calcul avec constructeurs de types, un problème d'unification est dit 
{\it élémentaire au niveau des types} si  pour toute variable 
$Qx:T$ déclarée dans $\Gamma$, le terme $T$ est l'un des termes
$Prop$, $Prop \ra Prop$, $Prop \ra Prop \ra Prop$ ou 
$Prop \ra Prop \ra Prop \ra Prop$ et le type commun de $a$ et $b$ est $Prop$.
\end{fdefinition}

\begin{fproposition}
Dans un calcul quelconque du cube, il n'existe pas de méthode effective 
permettant de décider si un problème d'unification élémentaire au niveau 
des termes a une solution.

Dans un calcul avec constructeurs de type,
il n'existe pas de méthode effective permettant de décider si un
problème d'unification élémentaire au niveau des types a une solution.
\end{fproposition}

\begin{ftheorem}
Dans un calcul avec types dépendants, il n'existe pas de méthode effective
permettant de décider si un problème de filtrage dont toutes les variables
existentielles sont du troisième ordre au plus a une solution.
\end{ftheorem}

\preuve{Pour tout problème d'unification élémentaire au niveau
des termes $\langle \gamma, u_{1}, u_{2}\rangle $ on construit un  problème de filtrage
$\langle \Gamma,t_{1},t_{2}\rangle $ 
tel que tous les types des variables existentielles de $\Gamma$ 
soient du troisième ordre au plus et 
$\langle \gamma, u_{1}, u_{2}\rangle $ a une solution si et seulement si 
$\langle \Gamma,t_{1},t_{2}\rangle $ en a aussi une.
On pose:
$$\Gamma = \gamma[\fa z:U;\fa P:U \ra Prop;\fa c:(P~z); \fa d:(P~z);
\fa G:(P~z) \ra (P~z) \ra (P~z);$$
$$\ex f:(h:U \ra U)(P~(h~u_{1})) \ra (P~(h~u_{2}))]$$
$$t_{1} = (G~(f~[x:U]z~c)~(f~[x:U]z~d))~~~~~~~~t_{2} = (G~c~d)$$

S'il existe une substitution $\tau$ telle que
$\tau u_{1} = \tau u_{2}$ alors on pose :
$$\sigma =\tau \cup 
\{\langle f,[],[x_{1}:U \ra U][x_{2}:(P~(x_{1}~(\tau u_{1})))]x_{2}\rangle \}$$
La substitution $\sigma$ est bien-typée dans $\Gamma$ et est solution du
problème de filtrage $\sigma t_{1} = t_{2}$.

Réciproquement, si il existe  une substitution $\sigma$ bien-typée dans 
$\Gamma$ telle que
$\sigma t_{1} = t_{2}$, alors $\sigma$ est bien-typée dans $\gamma$ et
on montre que $\sigma u_{1} = \sigma u_{2}$.
Soit :
$$\Delta = \Gamma 
[\fa x_{1}:U \ra U; \fa x_{2}:(P~(x_{1}~(\sigma u_{1})))]$$ 
$$v = ((\sigma f)~x_{1}~x_{2}):(P~(x_{1}~(\sigma u_{2})))$$
L'équation $\sigma t_{1} = t_{2}$ est équivalente au système :
$$v[x_{1} \la [x:U]z, x_{2} \la c] = c~~~~v[x_{1} \la [x:U]z, x_{2} \la d] = d$$
Le terme $v$ n'est pas une abstraction ni un produit parce que son type est
$(P~(x_{1}~(\sigma u_{2})))$.
C'est donc un terme atomique $v = (x~r_{1}~...~r_{p})$ où $x$ est une
variable ou l'un des symboles $Prop$ et $Type$.
Le symbole $x$ est l'une des variables $x_{1}, x_{2}, c$ parce que
$v[x_{1} \la [x:U]z, x_{2} \la c] = c$.
Il est différent de $c$ parce que $v[x_{1} \la [x:U]z, x_{2} \la d] = d$.
Il est différent de $x_{1}$ parce que le type de $v$ est différent de 
$U \ra U$ et $U$.
Donc $x = x_{2}$. Comme le type de $x_{2}$ est atomique $p = 0$ et $v = x_{2}$.
Les termes $v$ et $x_{2}$ ont donc m\^eme type et on en déduit :
$$\sigma u_{1} = \sigma u_{2}$$}

\begin{ftheorem}
Dans un calcul avec constructeurs de types, il n'existe pas de méthode
effective
permettant de décider si un problème de filtrage dont toutes les variables
existentielles sont du troisième ordre au plus a une solution.
\end{ftheorem}

\preuve{Pour tout problème d'unification élémentaire
au niveau des types
$\langle \gamma, u_{1}, u_{2}\rangle $ on construit un  problème de filtrage
$\langle \Gamma,t_{1},t_{2}\rangle $ 
tel que tous les types des variables existentielles de $\Gamma$ 
soient du troisième ordre au plus et 
$\langle \gamma, u_{1}, u_{2}\rangle $ a une solution si et seulement si 
$\langle \Gamma,t_{1},t_{2}\rangle $ en a aussi une.
On pose :
$$\Gamma = \gamma[\fa Z:Prop; \fa c:Z; \fa d:Z; \fa G:Z \ra Z \ra Z;
\ex f:(h:Prop \ra Prop)(h~u_{1}) \ra (h~u_{2})]$$
$$t_{1} = (G~(f~[X:Prop]Z~c)~(f~[X:Prop]Z~d))~~~~~~~~t_{2} = (G~c~d)$$

S'il existe une substitution $\tau$ telle que
$\tau u_{1} = \tau u_{2}$ alors on pose :
$$\sigma =\tau \cup 
\{\langle f,[],[x_{1}:Prop \ra Prop][x_{2}:(x_{1}~(\tau u_{1}))]x_{2}\rangle \}$$
La substitution $\sigma$ est bien-typée dans $\Gamma$ et est solution du
problème de filtrage $\sigma t_{1} = t_{2}$.

Réciproquement, si il existe  une substitution $\sigma$ bien-typée dans 
$\Gamma$ telle que
$\sigma t_{1} = t_{2}$, alors $\sigma$ est bien-typée dans $\gamma$ et
on montre que $\sigma u_{1} = \sigma u_{2}$.
Soit :
$$\Delta = 
\Gamma[\fa x_{1}:Prop \ra Prop; \fa x_{2}:(x_{1}~(\sigma u_{1}))]$$ 

$$v = ((\sigma f)~x_{1}~x_{2}):(x_{1}~(\sigma u_{2}))$$
L'équation $\sigma t_{1} = t_{2}$ est équivalente au système :
$$v[x_{1} \la [X:Prop]Z, x_{2} \la c] = c~~~~
v[x_{1} \la [X:Prop]Z, x_{2} \la d] = d$$
Le terme $v$ n'est pas une abstraction ni un produit parce que son type est
$(x_{1}~(\sigma u_{2}))$.
C'est donc un terme atomique $v = (x~r_{1}~...~r_{p})$ où $x$ est une
variable ou l'un des symboles $Prop$ et $Type$.
Le symbole $x$ est l'une des variables $x_{1}, x_{2}, c$ parce que
$v[x_{1} \la [X:Prop]Z, x_{2} \la c] = c$.
Il est différent de $c$ parce que $v[x_{1} \la [X:Prop]Z, x_{2} \la d] = d$.
Il est différent de $x_{1}$ parce que le type de $v$ est différent de 
$Prop \ra Prop$ et $Prop$.
Donc $x = x_{2}$. Comme le type de $x_{2}$ est atomique $p = 0$ et $v = x_{2}$.
Les termes $v$ et $x_{2}$ ont donc m\^eme type et on en déduit :
$$\sigma u_{1} = \sigma u_{2}$$}

\section*{Conclusion}

Le filtrage est donc indécidable dès le troisième ordre dans les calculs
$\lambda P (LF)$, $\lambda P2$, $\lambda \underline{\omega}$, 
$\lambda \omega (F_{\omega})$,
$\lambda P \underline{\omega}$ et $\lambda P \omega (CoC)$.
Restent ouverts le problème classique de la décidabilité du filtrage dans 
le $\lambda$-calcul simplement typé (problème conjecturé décidable dans 
\cite{Huet76}) et celui de la décidabilité du filtrage dans le système 
$\lambda 2 (F)$.

{\small {Auteur: Gilles Dowek, INRIA, Domaine de Voluceau-Rocquencourt, 
B.P. 105, 78153 Le Chesnay Cedex.}}

\newpage

\begin{center}{\Large{\bf Erratum au compte rendu :

L'Indécidabilité du Filtrage du Troisième Ordre dans les Calculs
avec Types Dépendants ou Constructeurs de Types}}\\[20pt]
{\bf Gilles Dowek}\\[10pt]
\end{center}

\begin{footnotesize}
 
Dans le compte rendu ``L'indécidabilité du filtrage du troisième ordre
dans les calculs avec types dépendents et constructeurs de types''
(I, 312, 12, 1991, pp. 951-956), le second théorème (p. 956) est erroné.

\bigskip

In the note ``The undecidability of third order pattern matching in calculi
with dependent types or types constructors'' (I, 312, 12, 1991,
pp. 951-956), the second theorem (p. 956) is invalid.

\end{footnotesize}

\bigskip

M. Bezem et J. Springintveld ont trouvé une
erreur dans la démonstration du second théorème de ce compte rendu.
Le terme $(h:Prop \rightarrow Prop)(h~u_{1}) \rightarrow (h~u_{2})$
n'est bien typé que dans un calcul comprenant des constructeurs de
types et des types polymorphes. Ce résultat d'indécidabilité ne 
s'applique donc pas aux calculs $\lambda \underline{\omega}$ ni 
$\lambda P \underline{\omega}$.
Par ailleurs, l'ordre de ce type n'est pas $3$ mais $\infty$. 
Le résultat démontré est donc l'indécidabilité du filtrage (et
non du filtrage du troisième ordre) dans les calculs avec types polymorphes
et constructeurs de types (et non dans les calculs avec constructeurs
de types). 

M. Bezem et J. Springintveld ont montré qu'il est possible d'adapter
la démonstration de façon à obtenir l'indécidabilité du filtrage du
quatrième ordre dans les calculs avec types polymorphes et
constructeurs de types, en considérant le problème $\langle \Gamma, t_{1}, t_{2}\rangle $ où
$$\Gamma = [\forall P:Prop \rightarrow Prop; \forall Z:Prop; \forall c:(P~Z); \forall d:(P~Z);$$
$$\forall G:(P~Z) \rightarrow (P~Z) \rightarrow (P~Z); 
\exists f:(h:Prop \rightarrow Prop) (P~(h~u_{1})) \rightarrow (P~(h~u_{2}))]$$
$$t_{1} = (G~(f~[X:Prop]Z~c)~(f~[X:Prop]Z~d))$$
$$t_{2} = (G~c~d)$$
J. Springintveld a par ailleurs montré la décidabilité du
filtrage du troisième ordre dans le système $\lambda
\underline{\omega}$, mais l'indécidabilité de ce problème
quand on impose aux solutions d'être closes. 
La décidabilité du filtrage dans le système
$\lambda \underline{\omega}$ ainsi que celle du filtrage du troisième ordre
dans $\lambda \omega$ restent ouvertes. 

Le premier théorème du compte rendu, concernant les calculs avec
types dépendants n'est pas affecté.

\end{document}